\documentclass[arxiv,times,nonumber]{sae}

\input{00-preambles}

\PaperTitle{Identification and Verification of Attack-Tree Threat Models in Connected Vehicles}%
\AddAuthor{Masoud Ebrahimi, Christoph Striessnig}{Graz University of Technology, Graz, Austria}
\AddAuthor{Joaquim Castella Triginer}{Virtual Vehicle Research GmbH, Graz, Austria}
\AddAuthor{Christoph Schmittner}{Austrian Institute of Technology, Vienna, Austria}

\begin{document}
\raggedbottom
\maketitle
\thispagestyle{plain}
\pagestyle{plain}
\begin{abstract}
As a result of the ever-increasing application of cyber-physical components in the automotive industry, cybersecurity has become an urgent topic.
Adapting technologies and communication protocols like Ethernet and WiFi in connected vehicles yields many attack scenarios.
Consequently, ISO/SAE~21434 and UN~R155~(2021) define a standard and regulatory framework for automotive cybersecurity.
Both documents follow a risk management-based approach and require a threat modeling methodology for risk analysis and identification.
Such a threat modeling methodology must conform to the Threat Analysis and Risk Assessment (TARA) framework of ISO/SAE~21434.
Conversely, existing threat modeling methods enumerate isolated threats disregarding the vehicle's design and connections.
Consequently, they neglect the role of attack paths from a vehicle's interfaces to its assets. In other words, they are missing the TARA work products, e.g., attack paths compromising assets or feasibility and impact ratings.
We propose a threat modeling methodology to construct attack paths by identifying, sequencing, and connecting vulnerabilities from a valid attack surface to an asset.
Initially, we transform cybersecurity guidelines to attack trees, and then we use their formal interpretations to assess the vehicle's design.
This workflow yields compositional construction of attack paths along with the required TARA work products (e.g., attack paths, feasibility, and impact).
More importantly, we can apply the workflow iteratively in the context of connected vehicles to ensure design conformity, privacy, and cybersecurity.
Finally, to show the complexity and the importance of preemptive threat identification and risk analysis in the automotive industry, we evaluate the presented model-based approach in a connected vehicle testing platform, SPIDER.
\end{abstract}
\section{Introduction}\label{sections:introduction}

Automotive cybersecurity has become an urgent topic with the change in the automotive industry from purely mechanical to cyber-physical systems.
Alongside numerous advantages, standardization efforts will lead to widely established cyber-physical components in the automotive domain, which automatically yield numerous existing attack tools and potential attack scenarios compared to proprietary or automotive-specific protocols.
An example of such a scenario is the automotive Ethernet, one of the preferred communication protocols in connected vehicles. Automotive Ethernet has the potential to address many security-related challenges inherent to some of the automotive communication protocols. On the other hand, since Ethernet and its adjacent technologies are reusable in a broader set of application domains, it is of interest to a more extensive set of security experts, making it more susceptible to cybersecurity threats.

Indeed, the evolution of vehicular networks towards cyber-physical components and communication protocols like automotive Ethernet increases the urgency of automotive cyber-security.
This urgency is mainly because exploiting cybersecurity threats in the automotive industry now requires less specific knowledge of the automotive architectures.
Meanwhile, many cybersecurity threats in the automotive industry emanate from faulty system design as cyber-physical components and protocols like Ethernet and adjacent technologies implement many cybersecurity goals.
Consequently, \iso \cite{iso21434} and \un (2021) \cite{unece155} define a standard and regulatory framework for approaching automotive cybersecurity.
Both documents follow a risk management-based approach, with the crucial first step of risk analysis and identification.
Therefore, the automotive industry must implement a rigorous threat modeling methodology following the Threat Analysis and Risk Assessment (TARA) framework from \iso.
Cyber-threat management frameworks are fundamental to protecting vehicles from cyber-attacks and performing an accurate risk assessment.

\rev{
A threat modeling approach is suitable for automotive cybersecurity if it is \textit{repeatable}, \textit{traceable}, and \textit{actionable}.
Repeatable threat modeling ensures different experts arrive at the same threat list and changes in system architecture lead to consistent changes in threat identification results.
A threat modeling approach provides traceability if it is possible to trace and determine which part of the system or configuration was the reason for any given threat.
Finally, a vehicle has around 140-200 ECUs, often utilizing older technologies;
thus, an actionable automotive cybersecurity concept addresses potential attacks at the interfaces and can later result in system-level mitigation strategies.
}





Contemporary threat modeling methodologies in the automotive industry do not provide threat modeling work products that conform to the TARA requirements, \eg, attack paths compromising system assets or feasibility and impact ratings.
To determine why current threat management methods do not provide such capabilities, we must recap the definition of an attack path: a sequence of steps from a valid attack surface leading to an asset in the system.
Attack paths are essential for determining threat severity value calculated from the feasibility and impact ratings of the threats in offensive and defensive security scenarios.
Unfortunately, in the automotive industry, current threat modeling methodologies disregard the concept of attack paths, resulting in work products that lack meaningful information for risk assessment in the context of the design under test.
This shortcoming is mainly because these methodologies enumerate isolated threats overlooking the system's design and disregarding the role of attack paths from the system's interfaces to its assets in computing potential threats' feasibility and impact ratings.

We propose a threat modeling methodology in conjunction with a threat modeling tool
called ThreatGet.
We define our methodology based on a formal approach that identifies and connects vulnerabilities to determine how to differentiate between the more extensive set of potential but isolated vulnerabilities and the smaller set of attack combinations and attack paths.
We assess the system design through formal interpretations of cybersecurity guidelines that we transformed into attack trees to answer this question.
This transformation leads to a compositional construction of detailed attack scenarios.
Hereafter, we can create threat rules containing the specific system design's TARA requirements (\eg, attack paths, feasibility, and impact ratings).
\rev{%
Since our approach is a model-based one, it ensures traceability from identified threats through attack paths.
Moreover, the construction of attack paths yields repeatability meaning the impact of model refinements is directly visible in the verification process.
Finally, the workflow of the proposed automated model-based methodology is based on \iso and therefore compliant with the standard, including the production of the required work products.
Attack Flow based analyses can also guide cybersecurity concepts and architecture design, by placing security measures at critical points along the attack paths.}
More importantly, we evaluate the presented model-based approach in a connected vehicle testing platform, SPIDER.
This example shows the results of applying the approach to an existing cyber-physical vehicle testing and the complexity of risk analysis and
identification for the complete vehicle architecture.
It demonstrates how we can pre-emptively avoid numerous threats involving popular technologies like Ethernet.
Finally, it aligns the results according to TARA activities defined in ISO/SAE~21434 and complies with the UN~R155.

The rest of this paper reads as follows.
\cref{sections:stateoftheart} reviews the state-of-the-art literature in automotive cybersecurity.
\cref{sections:ISO_SAE_21434_framework} describes the \iso framework and its relevance in the context of our contributions.
\cref{sections:methodology} describes the proposed methodology to implement the \iso framework described in the preceding section.
\cref{sections:Alignment} studies how the proposed methodology aligns with the requirements of \iso Standards and its TARA work products.
\cref{sections:UseCase} applies the proposed methodology to a real-world system whose purpose is to provide a testing platform for connected vehicles.
Finally, \cref{sections:conclusion} concludes this manuscript.
\section{State of the Art}\label{sections:stateoftheart}

\subsection{Cooperative Intelligent Transport Systems}

The automotive domain is transforming from disconnected and isolated vehicles, which only share a common road infrastructure, into a network of vehicles that share sensors and information.
The desire to achieve increased efficiency and safety drives this transformation \cite{ dajsuren2019automotive}. 
Like the change from mechanical controlled automotive systems towards electronically and software-controlled systems,
this transformation also begins with a singular vehicle but continues with softening the boundaries of vehicles.

Cooperative Intelligent Transportation System (C-ITS) is a transportation system in which infrastructure and vehicles share information and react accordingly. 
These connected capabilities range from simple services like information on traffic light phases to planning the approach towards warning vulnerable road users (cyclists and pedestrians) outside the field of view of single vehicles and even automated braking and route optimization. 

In the past C-ITS was mainly a topic for research projects and demonstrations of potential applications. Issues were the availability of communication infrastructure (\eg, roadside equipment), communication and processing equipment in the vehicle (onboard equipment), and topics of interoperability and agreement on available services. Starting in 2014, the European Commission increased its interest in C-ITS for reducing traffic fatalities. 
Consequently, C-ITS services were defined, and roadside equipment was installed along main traffic corridors in Europe \cite{lu2018c, kotsi2020overview}. In addition, since 2019, C-ITS has been available as standard equipment in series production vehicles \cite{sjoberg2020automotive}. 
With this, C-ITS will move from research toward daily life.

A working C-ITS system has the potential to increase efficiency and prevent accidents.
A C-ITS system is identified by sharing sensor information to create a shared awareness of traffic conditions and potential hazards. 
C-ITS services are of different service levels. 
Day 1 services are services in which the control of the vehicle is still in the hands of the human driver, \eg, The human driver receives the C-ITS information, but (s)he still handles interpretation and reaction. 
Day 1.5 services are currently the highest available level for series vehicles, and here some automation (emergency braking, self-parking) is considered \cite{lu2018c}. 
Day 2 services are still restricted to research and demonstration, and here partial automation of the vehicle is assumed, and the vehicle partially interprets and reacts to C-ITS information. 

Already the Day 1 services open up the communication networks inside vehicles and create a more extensive overlaying communication system. 
With Day 1 services, there is the potential for isolation between active components in the vehicle (like braking, steering, engine control) and outside exposed sensors and communication systems. 
With Day 1.5 and Day 2, such isolation will no longer work. 
The system must distribute the communicated information between active components. 
With this, ensuring security and a sufficient level of trust is of utmost importance.

\subsection{Automotive Security Analysis}
Risk management can ensure levels of trust.
Therefore, risk management was at the core of automotive security, starting from results for research projects (EVITA, HEAVENS) \cite{henniger2009securing, macher2016review} towards the first published SAE guidebook (SA J3061) \cite{schmittner2016using} towards the first international standard (\iso) \cite{macher2020iso} and regulation (\un) \cite{schmittner2020preliminary}.

Initial approaches from EVITA or HEAVENS and SAE J3061 mainly focused on single vehicles or sensitive components inside a vehicle. 
Such approaches, therefore, often neglected potential attack paths and focused on direct threats to components or a small combination of components. 
With the transition from isolated vehicles to C-ITS, there is a need to extend this focus and secure an interconnected network of vehicles and infrastructure. Consequently, \iso prescribed the usage of attack path analysis. 
With attack path analysis and attack tree analysis, cybersecurity experts can detect endangered attack surfaces and tailor risk management measures. 

With this in mind, approaches like threat modeling are no longer sufficient, and a comprehensive methodology is needed. Therefore there is ongoing work to extend established methodologies like HEAVENS towards a form where they are compliant with \iso and able to cover complex, and large C-ITS systems \cite{lautenbach2021proposing}. 

A major challenge in real-world systems is the complexity of architecture, variety of attack surfaces, and the number of assets that consequently result in an explosion in the number of attack paths. 
We present here an approach that utilizes attack trees and formulates them as reusable anti-patterns, which one can merge to detect commonalities between multiple attack scenarios. 
This capability enables cybersecurity experts to handle high-complexity scenarios with multiple attack trees and optimize risk management measures; that is why we used the fundamental security analysis of \iso as a reference point. 
\section{\iso Framework}\label{sections:ISO_SAE_21434_framework}

The impact of cybersecurity in the future of intelligent and connected vehicles requires a common understanding of the cybersecurity perspective in engineering electrical and electronic (E/E) systems within road vehicles. 
The standard for automotive cybersecurity \iso \cite{iso21434} specifies guidelines for cybersecurity risk management and enables organizations to define cybersecurity policies and processes and foster a cybersecurity culture. 
The standard specifies engineering requirements for cybersecurity risk management regarding the concept, product development, production, operation, maintenance, and decommissioning. Nevertheless, we are primarily interested in TARA methods for cybersecurity risk management and mainly focus on the Concept and Product Development phases.

\subsection{Item Definition}
The item definition declares the scope of analysis of a component or set of components that implement a function at the vehicle level. It contains the description of the item boundary, the item functions, and the preliminary architecture and defines the operational environment and its interactions in the context of cybersecurity. 

\paragraph{System Modeling:}
The item definition compiles the required information for the system modeling. This representation of the item is the starting point for our methodology. 
The system can be in its design stage with incomplete implementation details or a fully developed vehicle. 
As ThreatGet aims for high-level threat analysis, the proposed methodology can and should be applied in the earliest design and development stages. 
Nevertheless, we will show later that analyzing a fully developed system is possible. 

\subsection{Threat Analysis and Risk Assessment}
\label{sections:ISO_SAE_21434_framework:TARA}
The TARA framework provide a threat-based approach to help identify, assess, prioritize, and control cybersecurity risks. The TARA comprises two processes that result in impact and feasibility ratings, respectively. 
The first process starts by defining the valuable assets with the compromised cybersecurity properties and continues by describing the related damage scenario that describes the impact of a compromised asset and evaluates an impact rating. 
The second process continues by deriving the potential threats that could impact an asset leading to a damage scenario. 
This step defines an attacker's attack paths to reach and compromise an asset and assigns a feasibility rating to the attack path. 
Combining the impact and feasibility ratings creates a risk determination for each threat. 
The resulting risk is treated in one of the following options: avoidance, reduction, sharing, or retainment. Risk reductions lead to cybersecurity goals, and sharing or retainment leads to cybersecurity claims.%

\paragraph{Asset Identification:} 
An asset is valuable for the stakeholder and a worthwhile target for attackers. 
An asset can be a safety-critical function or confidential data. 
A compromised asset leads to a damage scenario. 
Asset identification in the system is the initial step in TARA methods and influences the system modeling analysis. 

\paragraph{Damage Scenario and Impact Rating:} 
A damage scenario specifies the effect of compromising an asset.
An impact rating assesses the potential damage to a system caused by a damage scenario. 
The rating ranges from negligible to severe in the following four categories (safety, financial, operational, and privacy). 

\paragraph{Threat Scenario:} 
While the damage scenario describes the potential impact, the threat scenario details how an attacker could compromise an asset. 
A threat scenario is a high-level description of what could cause the compromise of an asset. Multiple threat scenarios can lead to the same damage scenario, as there might exist different means to compromise the same asset.

\paragraph{Attack Path:} 
The attack path describes the detailed steps of a threat scenario. These steps focus on the attacker's point of view and include a sequence of actions that lead to the compromise of an asset. A threat scenario can have multiple attack paths.

\paragraph{Feasibility Rating:} 
The attack feasibility assesses the feasibility of an attack path, \ie, how feasible it is for the attacker to do each step and compromise the asset. 
One approach for calculating the feasibility is the CVSS exploit metric. The CVSS specification document \cite{team2015common} describes the four measurements used for this metric: attack vector (V), attack complexity (C), privileges required (P), and user interaction (U). By assessing these four measurements, we can calculate the exploitability metric. 

\paragraph{Risk Value:} 
A risk matrix mapping the impact and feasibility ratings to a severity rating determines the risk value.
The risk value indicates the overall severity of a threat.
TARA work products include detected threats and feasibility and impact ratings resulting in a risk value. 
However, the results include neither a description of the impacted asset nor a detailed damage scenario. 
Thus, we argue that the feasibility rating and risk value could be inaccurate without considering the attack path in applying TARA methods.

\subsection{Cybersecurity Goals}
Cybersecurity goals define requirements to reduce the risk imposed by a cybersecurity threat. 
The cybersecurity goals are identified after conducting the Threat Analysis and Risk Assessment. 

\subsection{Cybersecurity Concept}
Cybersecurity concepts define how to achieve cybersecurity goals. The concept includes technical and operational controls that prevent, detect, and monitor the compromise of the item. In addition, cybersecurity requirements are derived for each cybersecurity goal and allocated to the item following the derived technical and operational controls.

\subsection{Product Development}
Product Development is an iterative process that includes activities on decomposing requirements and creating system design and specifications. 
Most importantly, product development requires activities on the integration and verification of parts and their validation. 

The first activity breaks the item and its requirements into component and subcomponent specifications. 
We define cybersecurity specifications based on cybersecurity concepts and architectural design. 
Once the specifications are allocated into components and interfaces in the architectural design, we can implement the corresponding cyber-security designs. 

The second activity ensures that the implementation and integration of the cybersecurity designs comply with the defined cybersecurity specifications. This activity involves the definition of test cases and related test methods, including the application of extensive testing and the usage of test coverage as a metric for sufficiency. 

\paragraph{Validation:} 
Validation is the second part of the product development phase. The validations activities are applied at the vehicle level for the item considering the configuration for series productions. 
The validation confirms that the cybersecurity goals and claims are achieved and that no unreasonable risks remain.
The standard further handles the cybersecurity processes during production, operation, maintenance, and decommissioning, as well as continuous cybersecurity activities, like monitoring, event assessment, and vulnerability management. We will not discuss these steps here in detail as they are unrelated to our work. 
\iso only defines the general concept of cybersecurity management and only recommends different tools and processes for the actual analysis.

\section{Methodology}
\label{sections:methodology}
This section gives a set of considerations for formalizing and modeling a system using ThreatGet's extended \rev{D}ata-\rev{F}low \rev{D}iagram (DFD).
These considerations facilitate systematical specification of the system's weaknesses using a tree structure called an anti-pattern tree.
Given an anti-pattern tree, we study how to interpret it and derive threat rules.
Finally, we attach the ISO/SAE~21434~\cite{iso21434} work products to these threat rules using information collected throughout the process.

\begin{figure}[h!]
    \includegraphics[width=\columnwidth,page=1]{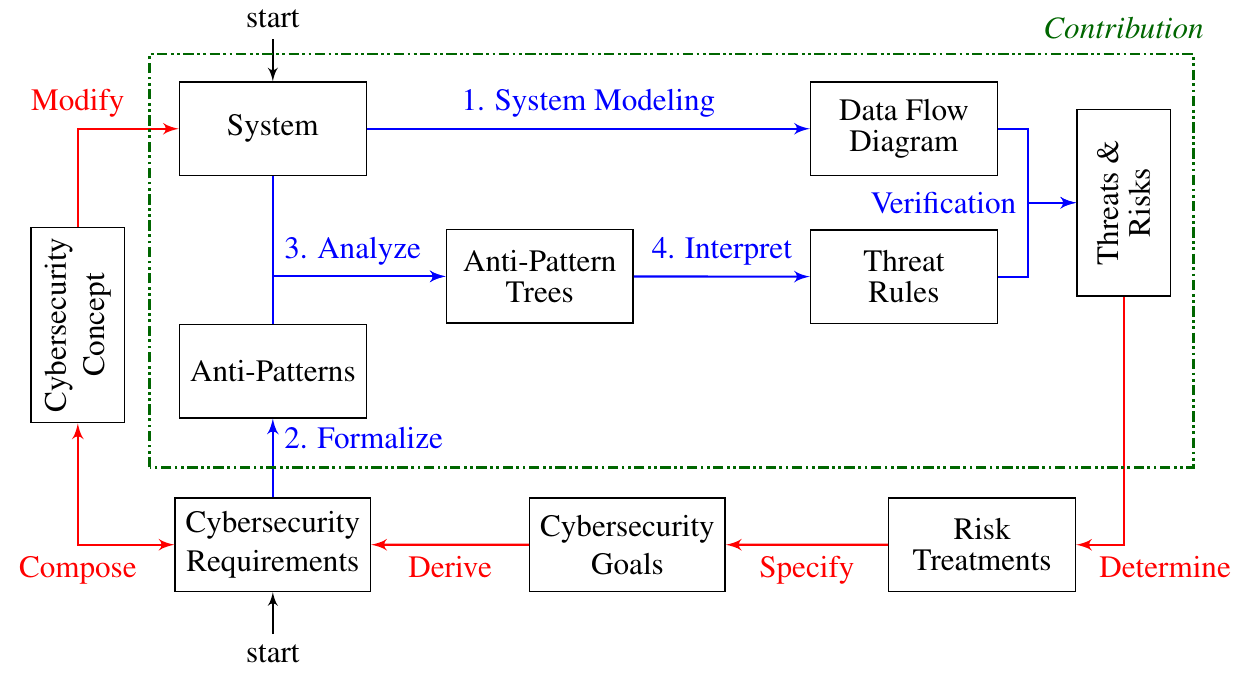}
    \caption[ThreatGet and \iso process.]{ThreatGet and \iso process.
    The \textcolor{blue}{blue} arrows are ThreatGet steps and the \textcolor{red}{red} arrows denote \iso steps.}
    \label{fig:main:process}
\end{figure}

\Cref{fig:main:process} demonstrate the proposed workflow, provides an overview of the combined process and highlights our \green{contribution}. We describe the complete workflow with an emphasis on our contribution; \ie the numbered \blue{processes} in \cref{fig:main:process}.

As it is a standard verification approach, there are two entry points in our workflow, \ie, the system under study and the cybersecurity requirements it must satisfy.
First, we describe the TARA methods and relevant work products, and then we describe the workflow.

\subsection{1.~System Modeling}
\label{subsection:systemmodeling}
The first step is to create an accurate model of the system.
ThreatGet's model is a DFD that defines the vehicles' elements and how they communicate with one another.

A ThreatGet DFD can comprise four component classes.
Each component might have a set of \textit{properties} describing its attributes or the countermeasures the system implements for the component.
Components have a unique \textit{type} that defines their symbols and properties.
Some types are within \textit{categories}.
A category defines a fixed set of properties, which all types within that category inherit.
A ThreatGet's DFD comprises the following components.
\begin{itemize}[leftmargin=*]
    \item \textbf{Elements} are component implementing a functionality.
    These are the core building blocks of DFDs; each data flow consists of elements that act upon the transferred data.
    Each element is of a unique type, \eg Sensor, Electronic Control Unit (ECU), \etc.
    Elements can have transitive, irreflexive, and asymmetric relations; \ie an element can contain one or more elements.
    We use the term \textit{child diagram} to refer to the contained elements.
    \item \textbf{Connectors} are directed links representing a data flow from a source to a destination element. Connecters are of wired and wireless types. Wired connectors include CAN, LIN, Ethernet, \etc, and wireless connectors include WiFi, Bluetooth, \etc.
    \item \textbf{Assets} mark valuable components held by elements and connectors, and if compromised, there is a negative impact on the system.
    According to the violated security properties, an asset can \rev{have} the following \rev{properties}: confidentiality, integrity, or availability.
    \item \textbf{Boundaries} enclose a collection of elements and are of two types: physical and logical.
    A physical boundary represents a car boundary, while a logical boundary represents a network segment.
    A boundary can contain other boundaries in a transitive, irreflexive, and asymmetric relation.
    Finally, a connector can cross a boundary.
\end{itemize}

To model a system in ThreatGet, we initially need to identify system components through the system's wiring and documentation.
Afterward, if possible, we match each component to a predefined element type, connector type, boundary type, or asset type using its attributes; otherwise, we will define a corresponding type for the component.
Of note, elements and connectors are structured categorically; thus, a component inherits all the properties in its category, and each anti-pattern defined for the category also affects the new component.
While categorizing each component, we define its properties by collecting relevant information through reviewing the system's documentation and interviewing developers, product owners, and end-users.
Once we discover all components, we must identify the vehicle's external interfaces.
Vehicles implement various communication methods with external components.
Either within proximity like WiFi, \rev{Vehicle-2-Everything (V2X)}, and Bluetooth, or globally through a mobile connection.
With all essential components and interfaces identified, we can model the system.

To model the system, we initially address each network segment of the vehicle within a logical boundary.
Depending on a segment's networking technology, we choose a fitting communication element whose type must conform to the network's connectors (\ie wired or wireless) and topology (\eg star or bus).
Afterward, we place the remaining elements within each segment and enclose all components within the vehicle with a physical car boundary.
Finally, we use wireless connectors to connect the external interfaces with their counterparts beyond the vehicle's physical boundaries.

Modeling a modern car with a hundred or more ECUs in a single diagram can be tedious and error-prone.
To model some aspects of the vehicle in more detail, we can add a child diagram to an element.
Child diagrams facilitate hierarchical modeling.
Hierarchical modeling improves a model's interpretability and provides an opportunity for a divide and conquer modeling paradigm.
Another use case for a hierarchical model is splitting a component into multiple elements.
For example, a control unit might consist of different software services with different cybersecurity requirements and technical restrictions.
Thus, we use a child diagram to model each service as a software element.

\begin{runexample}\label{runexample:dfd}
\rev{
Suppose a vehicle's ECU.
The ECU is connected to a WiFi interface by the bidirectional data flow.
It is also connected to a Lidar sensor, and a GPS sensor unidirectionally.
The DFD in \cref{fig:model:rex:dfd} illustrates the above elements.
}

\begin{figure}[H]
    \centering
    \includegraphics[page=2,scale=0.75]{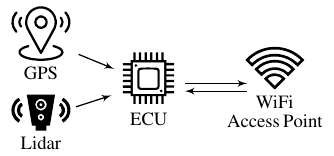}
    \caption{An example Data Flow Diagram in ThreatGet}
    \label{fig:model:rex:dfd}
\end{figure}
\end{runexample}

\subsection{2.~Formalize}
\label{sections:methodology:formalize}
The most crucial step in our workflow is to answer how to translate cybersecurity knowledge into a formal specification.
Before we start with the translation, we want to answer the following question: Instead of threat enumerations, should we translate cybersecurity statements into threat rules.
In the following, we make two orthogonal claims to prioritize cybersecurity statements over threat enumerations for the translation process:
\begin{enumerate}[leftmargin=*]
    \item Cybersecurity statements for secure systems define threats.
    \item A single cybersecurity statement covers multiple threats and, therefore, is more efficient to translate in the general case.
\end{enumerate}

\subsubsection{Anti-Patterns of Cybersecurity Statements}
To support our first claim, we have to define anti-patterns and constitute their relation with cybersecurity statements.
An \textit{anti-pattern} specifies a vulnerability or weakness that enables a threat.

\paragraph{Syntax:} ThreatGet defines its Domain Specific Language (DSL) to express anti-patterns.
The syntax consists of patterns and filters describing elements, connectors, assets, boundaries, and flows.
For more details on ThreatGet's anti-pattern syntax, refer to the work of ~\cite{DBLP:journals/corr/abs-2107-09986}.

\begin{example}[Flow Pattern]\label{example:flow:pattern}
A \flow pattern describes an arbitrary sequence of elements and connectors through the model.
It requires a source element pattern alongside a target element pattern and allows \texttt{CROSSES} and \texttt{INCLUDES} filters. The following anti-pattern specifies a flow whose source is an element of type \type{ElementTypeA} with a destination element of type \type{ElementTypeB} \st it includes an element of \type{ElementTypeC} with a \val{Value} for \field{Property}.
Finally, the flow shall not include a connector with a destination element of type \type{ElementTypeD} or must cross a boundary of type \type{BoundaryType}.
\end{example}

{\small
\begin{alltt}
\flow \{
    SOURCE \element: \type{"ElementTypeA"} &
    TARGET \element: \type{"ElementTypeB"} &
    INCLUDES \element: \type{"ElementTypeC"} \{
        \field{"Property"} == \val{"Value"}
    \} & \{
        INCLUDES NO \connector \{
            SOURCE \element & {TARGET} \element: \type{"ElementTypeD"}
        \} | CROSSES \boundary: \type{"BoundaryType"}
    \}
\}\end{alltt}}

As shown in \cref{example:flow:pattern}, the anti-pattern syntax can express complex combinations of the explained components.
Like filters, we can recursively combine anti-patterns using logical \AND and \OR to build more complex ones.

\paragraph{Semantics:}
The semantics of an anti-pattern constitute the model's correctness against the threat it describes.
The set of system components that matches an anti-pattern is a witness of the threat.
In other words, an anti-pattern specifies a system configuration whose existence yields a threat to the system.
In this work, we often refer to an anti-pattern as a weakness.
A \textit{weakness} is a design flaw that allows an attacker to take action and exploit the weakness.
Furthermore, if we use the term \textit{chain of weaknesses}, we refer to a combination of anti-patterns along a path.
We use the \flow pattern to express a chain of weaknesses in ThreatGet's DSL.
ThreatGet handles \flow patterns by applying a recursive depth-first-search algorithm.
In this work, we only discuss system modeling and creating threat rules for the analysis.
For more details about the syntax and analysis algorithm, refer to~\cite{DBLP:journals/corr/abs-2107-09986}.

\paragraph{From Cybersecurity Statements to Anti-Patterns:}
A collection of cybersecurity statements, whether they are mitigations (\eg \cite{unece155}) or technical practices (\eg \cite{enisaSecRep}), is a collection of logical propositions where each needs to be satisfied by the system should it be a secure system.
To formally state the above, suppose two cybersecurity statements specifying a secure system denoted by predicates $\phi$ and $\psi$ that must hold for all $x$, that is:
\begin{eqnarray*}
    \forall x : \phi(x) \land \psi(x)\;,
\end{eqnarray*}
where $x$ is a component of the system, like elements.
Alternatively, we can specify an insecure system by negating the predicate corresponding to a cybersecurity statement.
A negated cybersecurity statement is a \textit{threat rule} as it describes a possible system threat.
More formally,
\begin{eqnarray*}
\neg \left(\forall x : \phi(x) \land \psi(x)\right)
\Leftrightarrow & \exists x : \neg \left(\phi(x) \land \psi(x)\right) \\
\Leftrightarrow & \exists x : \neg \phi(x) \lor \neg \psi(x)\;.
\end{eqnarray*}

Threat rules are existentially quantified; meaning, there is an anti-pattern in ThreatGet's DSL to each threat rule that we can verify.
Therefore, given cybersecurity statements $\varphi(x)$ and $\psi(x)$, we can encode $\neg \phi(x)$ and $\neg \psi(x)$ as anti-patterns in ThreatGet's DSL to check whether the system is insecure.

\subsubsection{Threat Coverage of Cybersecurity Statements}
To underpin our second claim on threat coverage, we use the threat enumerations and mitigations provided by the UNECE Regulations No.155~\cite{unece155}.
We pick two threats from the list of threats to vehicle communication channels and state them and their mitigation in \cref{table:main:unece-threats}.
Both threats have the same mitigation, which is the usual case for most of the mitigation collection in the UNECE document.
Hence, should we translate the knowledge in the documents, it is generally more efficient to start with the cybersecurity statements; \ie mitigation.

\begin{table}[H]
    \footnotesize
    \centering
    \caption{Two threats from UNECE Regulations No.155~\cite{unece155}}
    \label{table:main:unece-threats}
    \begin{tabular}{p{8em} @{\hskip 1em} p{18em}}
        \toprule
        \textbf{Threat} & \textbf{Mitigation} \\
        \midrule
        \shortstack[l]{Spoofing by\\impersonation}
        &
        \shortstack[l]{The vehicle shall verify the authenticity\\and integrity of messages it receives}\\
        \midrule
        \shortstack[l]{MITM Attack /\\ Session Hijacking}
        &
        \shortstack[l]{The vehicle shall verify the authenticity\\and integrity of messages it receives}\\
        \bottomrule
    \end{tabular}
\end{table}

\begin{example}[TM-21]
Ensure that vulnerabilities and limitations of software dependencies, especially open source libraries, are mitigated or addressed in a risk assessment (TM-21).
Mitigation against such vulnerabilities is the ability to apply remote updates.
Therefore, our anti-pattern describes a third-party (or open-source) software element that does not provide update capabilities.

\begin{table}[H]
    \centering
    \footnotesize
    \caption{Anti-Patterns for TM-21}
    \label{table:main:tm21}
    \begin{tabular}{lp{17em}p{7em}}
        \toprule
        \textbf{ID} & \textbf{Pattern} & \textbf{Comment} \\
        \midrule
        $\upgamma_{0}$ &
\begin{alltt}
\element: \type{"Software"} \{
  \field{"License"} IN
    [\val{"open source"}, \val{"third party"}] &
  \field{"Updates"} NOT IN [\val{"remote"}, \val{"yes"}]
\}\end{alltt}%
        &
        \textbf{Targets:}\newline
        Integrity,\newline
        Authorization,\newline
        Confidentiality\xrule{7em}
        \textbf{Requires:} None
\\\bottomrule
\end{tabular}
\end{table}
\end{example}

\subsection{3.~Analyze}
The goal is to obtain threat rules that contain the impacted asset, a damage scenario, an impact rating, a threat scenario, an attack path, and a feasibility rating.
Consequently, verification results in almost all work products of the ISO/SAE~21434 TARA Methods\footnote{The results will not include risk treatment decisions.}.

Instead of focusing on threat rules, we introduce the concept of threat models to achieve the above goal.
A threat model is a tree of anti-patterns formalized by \textit{attack trees}.
The root of the tree is an asset of the system, while its leaves are the system's external interfaces.

To analyze, we first apply segmentation to divide the system into smaller segments.
Next, we identify assets within the segments and perform the anti-pattern activity per segment.
After that, we build anti-pattern trees that connect assets with their anti-patterns.
Finally, we merge the trees to create paths that cross multiple segments.

\subsubsection{Segmentation}
In the last decades, the automotive industry has experienced a transformation of the extension and complexity in the E/E vehicle architectures.
It is predictable that by considering interconnected intelligent vehicles and extending our cybersecurity analysis to the entire V2X infrastructure, the number of elements to be considered as targets of attackers increases exponentially.
In this context, segmentation represents a critical strategic activity that helps analysis proceed in an easy and manageable manner. In this paper, we strongly focused on vehicle systems.
Therefore, segmentation is focused on vehicle-level strategy, but this can be extended to more complex systems.

The goal is to divide the system into different segments. A segment is a part of a system that shares a level of privilege or access rights.
We use segments to collect elements through which an attacker either obtains certain access levels or obtains the potential to create a particular impact.
For example, the level of access to physical systems or the entry points to the system. The goal is to establish a hierarchical model of attack surfaces where the outermost segment contains initial attack surfaces, and the innermost segment contains the assets.

\begin{figure}[H]
    \centering
    \includegraphics[page=3,scale=0.8]{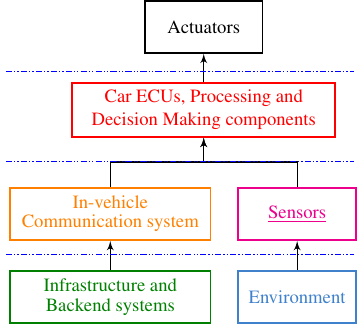}
    \caption{Reinterpretation of ENISA Smart Cars Reference Model~\cite{enisaSecRep}}
    \label{fig:model:enisa:tree}
\end{figure}

The segmentation strategy starts by analyzing the system boundaries. System boundaries distinguish the system from the context considering interactions in operational use. It provides information about the interfaces of those interactions with the environments. The second step continues with the system architecture transformation. Here we reinterpret the reference model by transforming it into a tree. Thus, it more closely represents the structure we use throughout our methodology. \cref{fig:model:enisa:tree} shows a reinterpretation of the High-Level Smart Cars Reference Model from the ENISA report \cite{enisaSecRep}. These data-flows represent the attack paths a remote attacker can take to obtain control of the vehicle. Finally, we perform the segmentation based on this hierarchical model. The system boundary indicates the initial entry points to the system, and the reinterpretation of the model the level of access to physical systems.

\subsubsection{Asset and Anti-Pattern Identification}
\label{sections:methodology:analyze:identification}

\paragraph{Assets:}~Asset identification consists of a process for identifying physical or logical item-related elements having either a perceived or actual value to the organization. Thus, assets are potential targets for an attacker of our system and mark valuable components.
According to the violated security properties, every asset can have a specific type, \eg, confidentiality, integrity, or availability.
Examples of assets are confidential data (private keys) or safety-critical data (messages that control cyber-physical actions).
This work generally relies on asset taxonomies to aid us with this task.
Specifically, we relied on the asset taxonomy from the ENISA report~\cite{enisaSecRep}.

\paragraph{Anti-Patterns:}
For every asset, there is a defined collection of anti-patterns. The main goal is to identify vulnerabilities for each asset and express them in anti-patterns through the translation of threats and cybersecurity statements. For this activity, collecting a catalog of various asset-related threats from multiple sources is strongly recommended. The refinement quality of such research will directly affect the number of anti-patterns and thus their quality. Again, this work uses the ENISA threats taxonomy~\cite{enisaSecRep} aligned with the previous asset identification. Nevertheless, we identify two criteria for each anti-pattern:
\begin{enumerate}[leftmargin=*]
    \item \textbf{Violated Cybersecurity Properties:} We identify one or more cybersecurity properties that the anti-pattern targets. The primary cyber-security properties are \textit{Confidentiality}, \textit{Integrity}, and \textit{Availability}. However, we use the extended STRIDE properties that also include \textit{Authentication}, \textit{Non-repudiation}, and \textit{Authorization}.
    \item \textbf{Entry Point:} An entry point enables an anti-pattern to be valid. It represents preliminary conditions to compromise some asset vulnerabilities in the system. For example, in our model, an attacker must first compromise the WiFi interface to gain access to the LAN.
\end{enumerate}
Each asset can have multiple anti-patterns, and each anti-pattern can have multiple entry points.

\begin{runexample}\label{example:WiFi}
\rev{
We continue our running example by identifying anti-patterns targeting the WiFi access point in the \cref{runexample:dfd}.
\Cref{table:main:ap1} gives two anti-patterns for this entry point interface $\iface_{\text{WiFi}}$.
The main task of the WiFi access point is user authorization.}
Thus, we create anti-patterns that mainly target the authorization property.
A valid weakness against the authorization function of $\iface_{\text{WiFi}}$ is that the element does not implement any authorization.
We include this simple example to ensure that the asset we should protect is disconnected from the actual element and that the element implements the countermeasures needed.
Thus, we define the anti-pattern $\upgamma_1$ as shown in \cref{table:main:ap1}.
Additionally, we define that $\upgamma_1$ targets authorization and confidentiality because the authorization mechanism of WPA2 also gives confidentiality to the data.
Joining an open WiFi requires no previous attacks; thus, it requires no entry point.
Another attack against $\iface_{\text{WiFi}}$ could be brute-forcing a short passphrase in WPA2.
The anti-pattern $\upgamma_2$ in \cref{table:main:ap1} describes a short passphrase that one can guess.

Finally, researchers discover design and implementation issues in the WiFi that help attackers bypass the authorization and confidentiality\footnote{https://www.krackattacks.com/}\footnote{https://www.fragattacks.com/}.
A third party is usually responsible for fixing these issues.
We cannot model such detailed attacks against our elements because the properties do not allow many details. We can model an attack that abuses outdated software issues by checking if the element supports updates. \cref{table:main:ap1} shows this pattern in $\upgamma_3$.

\begin{table}[H]
    \centering
    \footnotesize
    \caption{Anti-Patterns for the entry point interface $\iface_{\text{WiFi}}$}
    \label{table:main:ap1}
    \begin{tabular}{lp{17em}p{7em}}
        \toprule
        \textbf{ID} & \textbf{Pattern} & \textbf{Comment} \\
        \midrule
        $\upgamma_{1}$ &
\begin{alltt}
\element: \type{"WiFi Interface"} \{
    \field{"Authorization"} != \val{"yes"}
\}\end{alltt}
        &
        \textbf{Targets:}\newline
        Authorization,\newline
        Confidentiality\xrule{7em}
        \textbf{Requires:} None
        \\
        \midrule
        $\upgamma_{2}$ &
        \begin{alltt}
\element: \type{"WiFi Interface"} \{
    \field{"Key Length"} != \val{"long"}
\}\end{alltt}
        &
        \textbf{Targets:}\newline
        Authorization,\newline
        Confidentiality\xrule{7em}
        \textbf{Requires:} None
        \\
        \midrule
        $\upgamma_{3}$ &
        \begin{alltt}
\element: \type{"WiFi Interface"} \{
  \field{"License"} IN
    [\val{"open source"}, \val{"third party"}] &
  \field{"Updates"} NOT IN [\val{"remote"}, \val{"yes"}]
\}\end{alltt}
        &
        \textbf{Targets:}\newline
        Integrity,\newline
        Authorization,\newline
        Confidentiality\xrule{7em}
        \textbf{Requires:} None
        \\\bottomrule
    \end{tabular}
\end{table}
\end{runexample}

\begin{runexample}\label{example:sensors}
\rev{We continue our running example by identifying anti-patterns for Lidar and GPS interfaces in \cref{runexample:dfd}.
\cref{table:main:ap2} shows an anti-pattern for these sensor interfaces.
We can fuse data from multiple Lidar sensors to protect it against manipulation; thus, \ie $\upgamma_4$ checks if the Lidar sensor has no redundancy.}
Similarly, we must protect the integrity of the GPS interface through $\upgamma_5$.

\begin{table}[H]
    \centering
    \footnotesize
    \caption{Anti-Patterns for the sensor interface $\iface_{\text{sensor}}$}
    \label{table:main:ap2}
    \begin{tabular}{lp{16em}p{8em}}
        \toprule
        \textbf{ID} & \textbf{Pattern} & \textbf{Comment} \\
        \midrule
        $\upgamma_{4}$ &
\begin{alltt}\element: \type{"Lidar Sensor"} \{
  \field{"Redundancy"} NOT IN
  [\val{"double"}, \val{"triple"}]
\}\end{alltt}
        &
        \textbf{Targets:}\newline
        Integrity
        \xrule{8em}
        \textbf{Requires:} None
        \\ \midrule
        $\upgamma_{5}$ &
\begin{alltt}\element: \type{"GPS Interface"} \{
  \field{"Authentication"} != \val{"yes"} |
  \field{"Data security integrity"} NOT IN
    [\val{"encrypted"}, \val{"secure hash"}]
\}\end{alltt}
        &
        \textbf{Targets:}\newline
        Integrity,\newline
        Authentication,\newline
        Non-repudiation
        \xrule{8em}
        \textbf{Requires:} None
        \\
        \bottomrule
    \end{tabular}
\end{table}

\end{runexample}

\begin{runexample}
\Cref{table:main:ap3} gives anti-patterns for the ECU \rev{element in our running example}.
It deploys a web server for the operator to give commands.
The anti-pattern $\upgamma_6$ checks if the ECU applies both authentication and integrity to ensure communication security.
The exploitation of this weakness requires access to the WiFi and thus requires $\iface_{\text{WiFi}}$.
The second anti-pattern $\upgamma_7$ checks if the ECU validates the input it receives from the sensors.
An attack against this pattern requires an attack against the sensors denoted by $\iface_{\text{sensor}}$.

\begin{table}[t]
    \centering
    \caption{Anti-Patterns for the asset $\asset_{\text{ecu}}$}
    \label{table:main:ap3}
    \footnotesize
    \begin{tabular}{lp{16em}p{8em}}
        \toprule
        \textbf{ID} & \textbf{Pattern} & \textbf{Comment} \\
        \midrule
        $\upgamma_{6}$ &
\begin{alltt}
\element: \type{"ECU"} \{
  \field{"Authentication"} != \val{"yes"} |
  \field{"Data security integrity"} NOT IN
    [\val{"encrypted"}, \val{"secure hash"}]
\}\end{alltt}
        &
        \textbf{Targets:}\newline
        Integrity,\newline
        Authentication,\newline
        Non-repudiation
        \xrule{8em}
        \textbf{Requires:} $\iface_{\text{WiFi}}$
        \\
        \midrule
        $\upgamma_7$ &
        \begin{alltt}
\element: \type{"ECU"} \{
  \field{"Input Validation"} != \val{"yes"}
\}\end{alltt}
        &
        \textbf{Targets:}\newline
        Integrity
        \xrule{8em}
        \textbf{Requires:} $\iface_{\text{sensor}}$
        \\
        \bottomrule
    \end{tabular}
\end{table}

With the collection of assets and anti-patterns, we can move on to the next step of creating anti-pattern trees.

\end{runexample}

\subsubsection{Construction of Anti-Pattern Trees}

To formalize anti-pattern trees, we use attack trees.
Attack Trees are intuitive and straightforward structures to model adversarial actions. Schneier introduced the earliest version of attack trees in 1999 \cite{schneier1999attack}.
The trees describe how an attacker can reach their desired goal, like gaining access to a safe or someone's bank account.
The root of the tree represents the attacker's goal.
Each leaf is an attacker action, and the nodes in between represent subgoals.
To achieve the overall goal, an attacker must combine different actions to find a path to the tree's root.
A node can connect (or refine) its children by the conjunction (\AND) or disjunction (\OR).
These connections model the dependency between the children of a node and the node's goal.
If an attacker must achieve a goal with multiple requirements, then its children are connected by a conjunction.
If one action is sufficient, then a disjunction is used.
Jhawar \etal extended attack trees with sequential conjunction \SAND nodes \cite{DBLP:conf/sec/JhawarKMRT15}. The \SAND describes a time-based dependency of attacker actions.
An attacker must accomplish all different actions in a sequential order to satisfy \SAND connections.
A formal definition for an attack tree $\uptau$ is
\begin{eqnarray*}
    \uptau := \upgamma ~|~ \SAND(\uptau, \dots, \uptau)~|~ \AND(\uptau, \dots, \uptau)~|~ \OR(\uptau, \dots, \uptau)\;,
\end{eqnarray*}
where $\upgamma \in \upgamma$ and $\upgamma$ is the set of all attacker actions.

In the context of our methodology, we construct anti-pattern trees that consist of \rev{assets}, anti-patterns, and entry points.
We visualize the trees using logic gates to indicate their connections.
The \AND gates decorated with `$\to$' denote \SAND nodes.
An anti-pattern tree is a \SAND tree \st
\begin{eqnarray*}
    \uptau := \upgamma ~|~ \iface ~|~ \asset ~|~ \SAND(\uptau, \dots, \uptau)~|~ \AND(\uptau, \dots, \uptau)~|~ \OR(\uptau, \dots, \uptau)\;,
\end{eqnarray*}
where $\iface \in \mathcal{I}$ is an entry point in the set of possible interfaces $\mathcal{I}$, and $\asset \in \mathcal{A}$ is an asset in the set of all assets $\mathcal{A}$.

\begin{runexample}
We continue with our running example.
Based on \cref{table:main:ap1}, we construct a tree for entry point $\iface_{\text{WiFi}}$.
The tree's root is the entry point itself, and we connect the anti-patterns with an \OR node to the root.
We do the same for \cref{table:main:ap3}.
If the realization of an anti-pattern requires an entry point (\eg $\upgamma_6$), we connect that entry point with an \SAND node.
Conversely, if an anti-pattern is already realized (\eg $\upgamma_7$) and requires an entry point, we connect them with an \AND node.
This results in the trees shown in \cref{fig:main:ats:wifi,fig:main:ats:sensor,fig:main:ats:ecu}.
\end{runexample}

\subsubsection{Merging Anti-Pattern Trees}
To model complex threats, we merge the anti-pattern trees in two steps.
First, we select high-priority assets from the set of assets we obtained in \cref{sections:methodology:analyze:identification}.
The selection of high-priority assets should result in a subset of assets we wish to model\footnote{We do not define the criteria for selecting such assets in this work and leave this up to the users of our methodology.}.
Examples of high-priority assets are systems or user accounts with high privileges or safety-impacting functionality.
Second, we replace all entry points with the corresponding subtrees for each anti-pattern tree of the high-priority asset.
The scope of the trees at this point is quite limited and can be compared to the translated rules from \cref{sections:methodology:formalize}.
The difference is that we now have structures that allow us to combine them in more detailed patterns.

\begin{figure}[!th]
    \centering
    \begin{subfigure}[b]{0.4\columnwidth}
        \centering
        \includegraphics[page=5,scale=0.75]{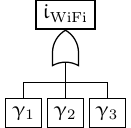}
        \setcounter{subfigure}{0}
        \caption{$\uptau_\text{WiFi}$}
        \label{fig:main:ats:wifi}
    \vspace{2mm}
    \end{subfigure}%
    \begin{subfigure}[b]{0.3\columnwidth}
        \centering
        \includegraphics[page=6,scale=0.75]{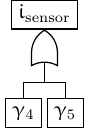}
        \setcounter{subfigure}{1}
        \caption{$\uptau_\text{sensor}$}
        \label{fig:main:ats:sensor}
    \vspace{2mm}
    \end{subfigure}
    \begin{subfigure}[b]{0.44\columnwidth}
        \centering
        \includegraphics[page=7,scale=0.75]{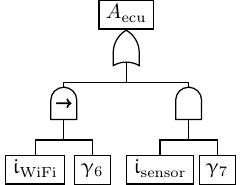}
        \setcounter{subfigure}{2}
        \caption{$\uptau_\text{ecu}$}
        \label{fig:main:ats:ecu}
    \end{subfigure}%
    \begin{subfigure}[b]{0.56\columnwidth}
        \centering
        \includegraphics[page=8,scale=0.75]{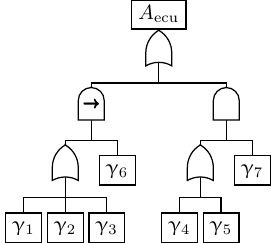}
        \setcounter{subfigure}{3}
        \caption{Merged anti-pattern tree of $\uptau_{\text{ecu}}$}
        \label{fig:main:at}
    \end{subfigure}%
    \caption{Anti-Pattern Trees $\uptau_{\text{WiFi}}$, $\uptau_{\text{sensor}}$, and $\uptau_{\text{ecu}}$}
    \label{fig:main:ats}
\end{figure}

\begin{runexample}
To illustrate this with our running example (depicted in \cref{fig:model:rex:dfd}), we choose $\asset_{\text{ecu}}$ as a high-priority asset.
The anti-pattern tree $\uptau_{\text{ecu}}$ (as shown in \cref{{fig:main:ats:ecu}}) has two entry points $\iface_{\text{WiFi}}$ and $\iface_{\text{sensor}}$.
Therefore, we merge the sub-trees $\iface_{\text{WiFi}}$ and $\iface_{\text{sensor}}$ into $\uptau_{\text{ecu}}$ by replacing the entry points with a connection to the respective sub-trees $\uptau_{\text{WiFi}}$ and $\uptau_{\text{sensor}}$.
See \cref{fig:main:at}.
\end{runexample}

By connecting the assets and entry points of different anti-pattern trees, we create a network to reason about paths that cross multiple segments.
These paths contain details about the initial attack vector, the complexity, and the impact.
We need these indicators to calculate impact and feasibility ratings and obtain the complete \iso work products and TARA Methods.

\subsection{4.~Interpret}
Given an anti-pattern tree, path enumeration results in threat rules describing sophisticated paths of weaknesses from system boundaries compromising system assets.
This is aligned with the description of attack paths in \iso work products.

\subsubsection{Path Interpretation}

In this section, we enumerate distinct anti-patterns of an anti-pattern tree.
To do this, we must cover the formal interpretation of \SAND attack trees by Jhawar \etal.~\cite{DBLP:conf/sec/JhawarKMRT15}.
They interpret \SAND trees using series-parallel (SP) graphs.
SP graphs are directed graphs that label their edges with the attacker's actions.
Each SP graph contains a single source and a single sink vertex.
The source has no incoming edges, while the sink has no outgoing edges.

The SP semantics of a tree $\uptau$ denoted by $\spgraph{\uptau} = \{G_1,\dots,G_n\}$ with $G_i$ being SP graphs for $i \in [1,n]$.
This set of SP graphs represents all the different paths through the attack tree that achieve its goal (\ie activate the root).
The \(\spgraph{\uptau}\) splits \(\uptau\) into distinct paths from leaf to root that obey the logic gates as follows.
In the graph, we interpret \SAND gates as sequential edges (denoted by $\Delta \cdot \nabla$) and \AND gates as parallel edges (denoted by $\Delta\,\|\,\nabla$).
Meanwhile, each \OR gate creates new graphs.
\Cref{fig:main:sp} depicts $\spgraph{\uptau_{\text{ecu}}}$ for the $\uptau_{\text{ecu}}$ shown in \cref{fig:main:at}.
We can write the SP graph interpretation of $\uptau_{\text{ecu}}$ as
\begin{eqnarray*}
    \llbracket \uptau_{\text{ecu}} \rrbracket_{\mathcal{SP}} =
    \{
        \upgamma_{1}\cdot\upgamma_{6},
        \upgamma_2\cdot\upgamma_{6},
        \upgamma_3\cdot\upgamma_{6},
        \upgamma_4\,\|\,\upgamma_{7},
        \upgamma_5\,\|\,\upgamma_{7}
    \}\;.
\end{eqnarray*}
\begin{figure}[H]
    \centering
    \begin{subfigure}[t]{0.225\columnwidth}
        \centering
        \includegraphics[page=9,scale=0.9]{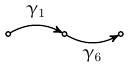}
    \end{subfigure}%
    \begin{subfigure}[t]{0.225\columnwidth}
        \centering
        \includegraphics[page=10,scale=0.9]{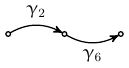}
    \end{subfigure}%
    \begin{subfigure}[t]{0.225\columnwidth}
        \centering
        \includegraphics[page=11,scale=0.9]{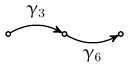}
    \end{subfigure}%
    \begin{subfigure}[t]{0.15\columnwidth}
        \centering
        \includegraphics[page=12,scale=0.9]{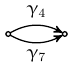}
    \end{subfigure}%
    \begin{subfigure}[t]{0.15\columnwidth}
        \centering
        \includegraphics[page=13,scale=0.9]{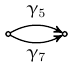}
    \end{subfigure}
    \caption{SP Graphs of the merged anti-pattern tree in \cref{fig:main:at}}
    \label{fig:main:sp}
\end{figure}
For each of the resulting graphs, we create a \flow pattern (covered in \cref{example:flow:pattern}) as a conjunction of the anti-patterns.
Of note, a \flow is semantically similar to an SP graph.
It has a single source and a single target node with some in-between.
The source vertex of the SP graph describes the source element of the \flow pattern, and the sink vertex describes the target element.
To include elements between source and sink, we use the \texttt{INCLUDES} filter in the \flow pattern.

Finally, we must address the transformation difference between sequential and parallel edges in the described process.
ThreatGet does not distinguish sequential dependencies of the SP graphs.
Thus, there is no difference between sequential and parallel anti-patterns.
We can use the source and target patterns in a \flow, but the elements in between can be of any order.
Nevertheless, we still use \SAND gates in tree modeling to provide a sequential building construct to the modeler.

\begin{runexample}\label{ex:main:fp1}
\rev{Continuing our running example, we can interpret the graph $\upgamma_1\cdot\upgamma_6$ to obtain the}
\flow pattern in \cref{table:running:rule} describing a chain of weaknesses with two anti-patterns from \cref{table:main:ap1,table:main:ap3}.
\end{runexample}

\subsubsection{Compiling Threat Rules}
\label{sec:main:compile}

This section addresses each ISO/SAE~21434 work product and explains how to obtain it.
Specifically, we complete the running example by providing the work products required by the \iso framework we described in \cref{sections:ISO_SAE_21434_framework}.
The following explains how to extract the information from our anti-pattern trees and \texttt{FLOW} patterns.
\cref{table:running:rule} shows the threat rule resulted from $\upgamma_1\cdot\upgamma_6$ in \cref{ex:main:fp1}.

\paragraph{Asset, Damage Scenario, and Impact Rating:}
We define each rule based on a targeted asset.
The asset is the root of the merged tree from which we derived the path of anti-patterns.
We derive a damage scenario for the asset.
The scenario is part of the threat rule and should give a high-level description of the trees' impact.
We derive the impact rating from the defined damage scenario using the safety mappings given in ANNEX~F~\cite{iso21434}.

\paragraph{Threat Scenario, Attack Path, and Feasibility Rating:}
A threat scenario is a more general description of the attack.
If we use threat enumerations to define our anti-patterns, we use them to define a threat scenario.
It is sufficient to summarize the attack path to define a threat scenario.
To define the attack path, we transform the path of weaknesses into a path of attacks that exploit those weaknesses.
We use the CVSS-based exploitability rating (E) we summarized in \cref{sections:ISO_SAE_21434_framework:TARA} to calculate our feasibility values.

\begin{table}[!t]
    \footnotesize
    \centering
    \caption{Threat-Rule for the SP graph $\upgamma_1\cdot\upgamma_6$}
    \label{table:running:rule}
    \begin{tabular}{p{5em}p{23.5em}}
        \toprule
        Title: &
        Compromise ECU through WiFi
        \\
        \midrule
        Asset: & High Level Controller \\
        \midrule
        Damage\newline
        Scenario: &
        A rear-end collision due to unintended high-speed braking.
        \\
        \midrule
        Attack\newline
        Path: &
        1.\,An attacker abuses an open WiFi access point to gain\newline access to the vehicle.\newline
        2.\,An attacker abuses the lack of authentication and\newline integrity to inject braking commands.
        \\
        \midrule
        Threat\newline
        Scenario: &
        The ECU issues braking commands to the low-level\newline controller due to unprotected communication channels.
        \\
        \midrule
        Threat~Type: & Tampering \\
        \midrule
        Feasibility: &
        Medium ($V\!\!=\!\textit{adjacent}, C\!\!=\!\textit{low}, P\!\!=\!\textit{none}, U\!\!=\!\textit{none}$)
        \\
        \midrule
        Impact: & Major -- Safety \\
        \midrule
        Rule: &
\begin{alltt}
\flow \{
  SOURCE \element: \type{"WiFi Interface"} \{
    \field{"Authorization"} != \val{"yes"}
  \} &
  TARGET \element: \type{"ECU"} \{
    \field{"Authentication"} != \val{"yes"} |
    \field{"Data security integrity"} NOT IN
      [\val{"encrypted"}, \val{"secure hash"}]
  \}
\}\end{alltt}\\
\bottomrule
\end{tabular}
\end{table}

\begin{runexample}
Since we need to be within range of the local area wireless network, the attack vector $V$ is \textit{adjacent}.
Since the WiFi is open, the attack complexity $C$ is \textit{low} and the privileges required $P$ is \textit{none}.
Finally, the required user interaction $U$ is \textit{none} as we do not rely on any user input.
This results in CVSS-exploitability $E$ equal to
\begin{center}
\vspace{0.5\parskip}
\footnotesize
\begin{tabular}{p{6em}p{6em}l}
    $V = 0.62\;,$&
    $C = 0.77\;,$&
    $E = 8.22 \times V \times C \times P \times U$\\[2mm]
    $P = 0.85\;,$&
    $U = 0.85\;,$&
    $~~~= 2.835\;,$
\end{tabular}%
\vspace{0.5\parskip}
\end{center}
which is within \textit{medium} exploitability feasibility \wrt ANNEX G \cite{iso21434}.
\end{runexample}

We define these work products from the collected information for each threat rule.
The final work product we address is the risk value.
As we stated in \cref{sections:ISO_SAE_21434_framework:TARA}, we derive the risk value from the impact and feasibility rating.
For each detected threat, ThreatGet automatically assigns the risk value \wrt the threat's impact and feasibility ratings.

Finally, we established a process for modeling the system and system-specific threat rules.
From here, the \iso~\cite{iso21434} TARA continues with the risk treatment decisions.
We collected a lot of information that will guide the risk treatments and help system engineers address the threats.
In the next chapter, we apply our methodology to the use case, providing further details for the steps and demonstrating how our methodology works in practice.
\begin{figure*}[!tbh]
    \centering
    \includegraphics[page=14,width=0.9\textwidth]{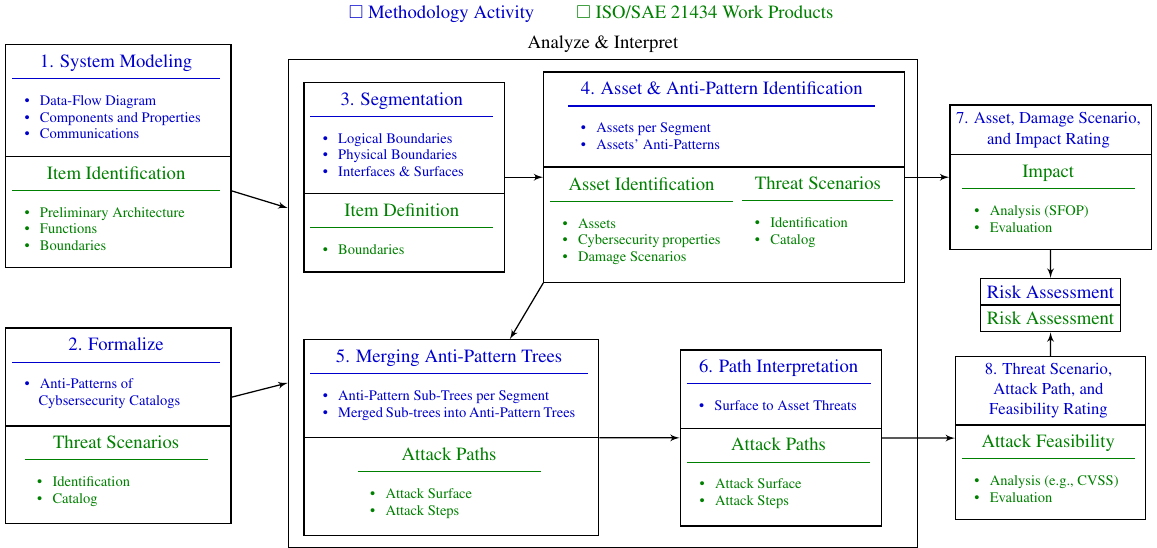}
    \caption{Methodology activities aligned with requirements/activities of the \iso}
    \label{figures:alignment}
\end{figure*}

\section{Aligning the methodology with \iso}\label{sections:Alignment}
\iso does not define a concrete approach or method for the TARA. Instead, a set of steps and results (\ie work products) with requirements is defined, which a compliant approach should produce. To demonstrate the compliance of our presented approach with \iso, \cref{figures:alignment} presents a mapping from \iso work products to the methodology's activities.

\subsection{Item definition}
Item definitions describe the preliminary architecture, known security measures, functions, and item boundaries.
In addition, it must include interactions with external systems.
System modeling \ref{subsection:systemmodeling} is the corresponding activity in the proposed methodology that results in the above work product.
A system model is the architecture of the items, interactions with external systems, and boundaries.

\subsection{Threat Scenarios}
A threat scenario describes a target asset, the compromised cyber-security properties\footnote{We are interested in extended STRIDE properties: Confidentiality, Integrity, Availability, Authentication, Non-repudiation, and Authorization.} of the asset, and a potential cause for the compromise, corresponding to the description of anti-patterns and documented assets in the system model in the proposed methodology.
The anti-pattern describes the targeted asset, property, and exploited weakness, \eg, cause for the compromise.

\subsection{Attack Path}
Attack paths describe how attacks might trigger a threat scenario. Based on the initial access of the attacker, a threat scenario describes all steps necessary to move from an attack surface to an asset contained in the threat scenario. Attack paths considered in isolation, \eg, formulation of a set of attack paths for each asset, starting from each attack surface and utilizing all potential paths through the system, can lead to many attack paths.
With Attack trees, the attack path is structured, and by merging and path interpretation, we achieve a comparable and reusable description of attack paths. Attack paths are the output of the steps from \textit{Construction of Anti-Pattern Trees} to \textit{Path Interpretation}.

\subsection{Risk Evaluation and Treatment}
\iso bases its risk evaluation on impact and attack feasibility.
The impact is based on the threat scenario, \eg, the violation of a cybersecurity property triggers a specific damage scenario.
While one can consider additional damage scenarios, the minimum required list includes damages to road users' safety, financial, operational, or privacy.
Safety damages correspond to the safety impact rating scheme from ISO 26262.
The proposed methodology accommodates this work product in \textit{Asset, Damage Scenario, and Impact Rating}.

Attack Feasibility is based on the Attack Path and utilizes the CVSS exploitability rating, which is included in the step of \textit{Threat Scenario, Attack Path, and Feasibility Rating}.
Here we can define CVSS ratings (or any other suitable rating) for the threat rules of an anti-pattern tree.
With both elements defined, a risk matrix determines the risk value based on impact and attack feasibility ratings.

\subsection{Cybersecurity Goals and derived Requirements}
\iso offers different approaches for the treatment of risks. If the selected treatment reduces the risk to an acceptable level, a cybersecurity goal that describes the risk reduction must be defined. A cybersecurity goal can thus be understood as a requirement to protect one or more assets against one or more threat scenarios or attack paths.
\section{Use Case}\label{sections:UseCase}

The proposed methodology addresses the evaluation of Intelligent Connected Vehicles in the context of cybersecurity.
To show the broad scope of the presented methodology, we evaluate that in a connected vehicle testing platform called SPIDER.
The system we analyze is the Smart Physical Demonstration and Evaluation Robot (SPIDER), which is a mobile Hardware-in-the-Loop (HiL) platform for testing perception systems and decision-making algorithms in real-world scenarios.
In addition, SPIDER is designed to handle V2X connectivity and integrates multiple remote services that interact with the surrounding environment.
For example, SPIDER external communication includes the Message Queuing Telemetry Transport (MQTT) messaging protocol that the automotive industry has embraced for V2X services.

\begin{figure*}[tbh]
    \centering
    \begin{subfigure}[b]{0.57\textwidth}
        \centering
        \includegraphics[page=15,width=\textwidth]{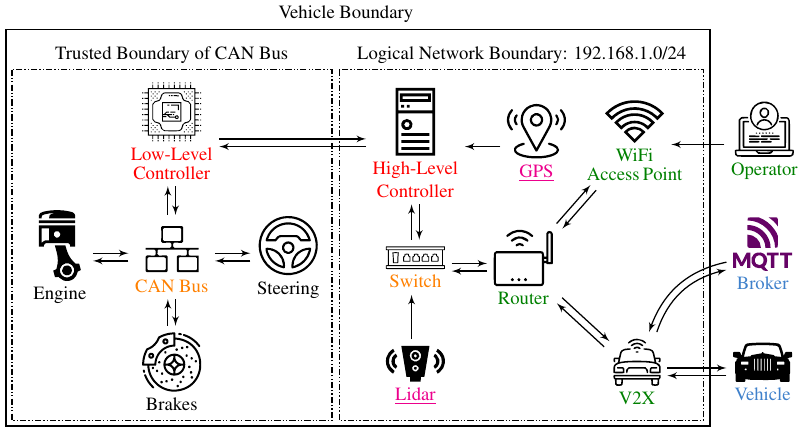}
        \caption{DFD of the SPIDER}
        \label{fig:model:spider:dfd}
    \end{subfigure}%
    \hfill
    \begin{subfigure}[b]{0.35\textwidth}
        \centering
        \includegraphics[page=16,width=\textwidth]{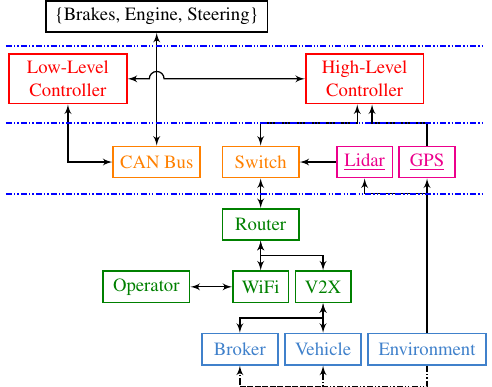}
        \caption{Tree representation of SPIDER's DFD}
        \label{fig:model:spider:tree}
    \end{subfigure}
    \caption{SPIDER High-Level System Model alongside its hierarchical tree representation according to the ENISA Reference Model in \cref{fig:model:enisa:tree}}
    \label{fig:model:spider}
\end{figure*}

\subsection{System Modelling}

The SPIDER architecture information is used to model the DFD in the ThreatGet.
The modeling starts with the elements in the in-vehicle network.
These are mainly interconnected using the Controller Area Network (CAN) and the Local Area Network (LAN).
\Cref{fig:model:spider:dfd} shows two logical boundaries alongside a physical car boundary around all elements.
These boundaries mark the segments of our system and visualize the components that can directly communicate.
What remains are the external elements with which our vehicle communicates.
An operator connects to the WiFi through the operator panel to control the vehicle.
The V2X element communicates with other vehicles and roadside infrastructure.
Finally, the LTE modem connects to a backend where an MQTT server broadcasts commands.

\subsection{Analysis}
To apply the proposed methodology to the SPIDER, we limit the scope of our threat model to remote attack vectors.
This section aims to find anti-patterns that represent remote attacks.
According to CVSS specification, this limits the attack vectors to Network and Adjacent.
This scope binds the attack paths to the network stack. The network attack vector describes possible exploits from a network adjacent to the target (\eg Bluetooth or WiFi) or possible remote exploits via the Internet.

\subsubsection{Segmentation}
We perform the segmentation strategy according to SPIDER architectural model, vehicle boundaries, and operational environment.
To continue with the segmentation, we first reinterpret SPIDER's DFD (shown in \Cref{fig:model:spider:dfd}) into a tree conforming to the reference model shown in \cref{fig:model:enisa:tree}. See \Cref{fig:model:spider:tree}.

It is easy to see that the vehicle boundaries and operational environment are the entry points that enable remote access to the system and can be potentially exploited.
We combine these entry points in the outermost segment.
We move inwards to the local area network to follow the idea of a remote attacker.
We defined the in-vehicle communication system as our next segment and moved up to the SPIDER components to reach our core segment, the actuators.

\paragraph{Environment:} Complete control of the automotive environment is almost impossible.
Many obstacles and dangers on the road can cause harm and put drivers at risk.
Traffic rules and signs help mitigate this risk and maintain road users and vehicle passengers in a safe environment.
However, some malicious actors may consciously or unconsciously carry out intentional vandalism, for example, a painted traffic sign or a driver exceeding the speed limit.

While devising a segmentation strategy, one must consider the environment is uncontrollable.
Therefore, it should identify the environment as the medium by which attackers can exploit vulnerabilities.
In other words, we must understand how our vehicle interacts with the outside world and under what conditions.
We assume the environment comprises all the external agents interacting with the V2X, such as other vehicles or infrastructure over which SPIDER has no control.

\paragraph{Sensors:}
To recognize the environment, modern cars and autonomous vehicles integrate from sixty to one-hundred sensors~\cite{DBLP:journals/corr/ISSN1530-437X}.
These sensors can estimate the distance between vehicle and object in the environment or recognize road signs or pedestrians.
The segmentation recognizes sensors that collect information from the environment and are susceptible to possible cyberattacks.
The SPIDER use case considers the cybersecurity risk of Lidar and GPS sensors.

\paragraph{Infrastructure, and Backend Servers:}
The SPIDER interacts with the outside world via the V2X interface.
V2X systems are part of the general IoT technologies that connect and exchange data with other devices and systems over the internet or other communications networks.
In this case, the V2X infrastructure operates with specific rules and protocols over the IoT environment within which malicious actors can also carry out deliberate attacks.
The V2X interface is connected to the SPIDER's Router and through a cellular interface, it connects to the IoT platforms.
For brevity, we focus strictly on the vehicle-level analysis of the V2X disregarding V2X interactions with systems outside SPIDER.

Next, we must consider the existing private infrastructure and backend servers responsible for vehicle-manufacturer communication.
In this case, Virtual Vehicle, the designer and manufacturer of the SPIDER, integrated a private network for remote control purposes.
\cref{fig:model:spider:tree} shows an operator that can remotely control certain vehicle parameters and collect data by establishing a connection with the WiFi access point.

\paragraph{In-vehicle Communication System:}
Since in-vehicle communication systems enable an attacker to access the network stack, it is a key segment to protect from attack vectors.
Protecting this asset can reduce the risk drastically.
For SPIDER, the Ethernet switch and the CAN bus are the two in-vehicle communications assets we must protect.

\paragraph{ECUs, Processing and Decision making Components:}
Various vehicular functions are distributed amongst many different ECUs.
For example, the Low-Level Controller provides basic vehicle driving functionalities, while the High-Level Controller performs ADAS functionalities, among others.
Depending on the level of refinement, vulnerabilities can be at the component level (\eg ECUs) but can also be analyzed at the subcomponent level (\eg services and programs that are running on ECUs).
Therefore, the segmentation can include levels of hierarchy at this stage, depending on the depth of the cybersecurity analysis. For brevity, we will not go beyond the component level.

\paragraph{Actuators:}
Since the actuators are ultimately responsible for the vehicle's movements, the tree reinterpretation of the SPIDER's model considers the actuators as the tree's root.
Finally, SPIDER is a vehicle test platform that we designed with four independent steering wheels that allow translational and rotational motions.
This provides huge actuator freedom during development but provides additional safety concerns while under cyber-attacks.

\subsubsection{Asset and Anti-Pattern Identification}
According to the methodology, we identify assets per segment and collect anti-patterns per asset.
The result is a set of tables (similar to \cref{table:main:ap1,table:main:ap3}), each of which describes an asset's list of anti-patterns.
Again for brevity reasons, we only focus on the infrastructure and backend servers segment that contains the remote attack vectors.
Items outside the vehicle boundaries, such as the operator, are not included.

\paragraph{WiFi Access Point:}
The WiFi access point creates a wireless network that is accessible from up to 300 meters through a WPA2 Pre-Shared Key (PSK).
\Cref{example:WiFi} already covers the anti-patterns' descriptions for entry point interface $\iface_{\text{WiFi}}$ and \cref{table:main:ap1} formalizes the anti-patterns $\upgamma_1$ to $\upgamma_3$ in ThreatGet's DSL.

\paragraph{Lidar and GPS:}
\Cref{example:sensors} already covers the anti-patterns' descriptions for Lidar and GPS interfaces.
Similarly, \cref{table:main:ap2} formalizes the anti-patterns $\upgamma_4$ and $\upgamma_5$ in ThreatGet's DSL.

\paragraph{V2X Interface:}
The V2X component uses Dedicated Short-Range Communication (DSRC) based on the IEEE 802.11 standards.
Thus, we again have different security measures on the link and application layers.
The link-layer security is similar to the WiFi access point.
However, on the application layer, the V2X component uses public-key infrastructure (PKI) based encryption to ensure the messages' authentication, confidentiality, and integrity.
This component could have different security measures on different network layers.
While we do not care about the security applied at the different layers in this high-level threat model, we must model the system correctly.

The only reason to distinguish between the layers is to ensure that the rule hits the element responsible for the security. To illustrate this point, we compare the V2X rules with the WiFi and cellular rules.
The question is: where the endpoint for the connection-oriented data flow is.
The V2X element creates an application-layer connection to the external elements it communicates with (other vehicles or infrastructure).
Thus, it is responsible for application-layer security.
Finally, in cooperative vehicles, the V2X communication contains safety-critical messages that must be available for all participants. Thus, the system needs to implement a DoS mitigation to withstand such attacks and adequately react to such messages; that is, anti-pattern $\upgamma_{11}$ in \cref{table:Av2x}.

\begin{table}[tbh!]
    \centering
    \footnotesize
    \caption{Anti-Patterns for V2X entry point interface $\iface_{\text{V2X}}$}
    \label{table:Av2x}
    \begin{tabular}{lp{17.5em}p{7.5em}}
        \toprule
        \textbf{ID} & \textbf{Pattern} & \textbf{Comment} \\
        \midrule
        $\upgamma_{9}$ &
\begin{alltt}
\element: \type{"V2X"} \{
    \field{"Authorization"} != \val{"yes"} |
    \field{"Data security integrity"} NOT IN
      [\val{"encrypted"}, \val{"secure hash"}]
\}\end{alltt}
        &
        \textbf{Targets:}\newline
        Integrity,\newline
        Authorization,\newline
        Non-repudiation\xrule{7.5em}\newline
        \textbf{Requires:} None
        \\
        \midrule
        $\upgamma_{10}$ &
        \begin{alltt}
\element: \type{"V2X"} \{
    \field{"Confidentiality"} != \val{"encrypted"}
\}\end{alltt}
        &
        \textbf{Targets:}\newline
        Confidentiality\xrule{7.5em}\newline
        \textbf{Requires:} None
        \\
        \midrule
        $\upgamma_{11}$ &
        \begin{alltt}
\element: \type{"V2X"} \{
    \field{"DoS Mitigation"} != \val{"yes"}
\}\end{alltt}
        &
        \textbf{Targets:}\newline
        Availability\xrule{7.5em}\newline
        \textbf{Requires:} None
        \\
        \midrule
        $\upgamma_{12}$ &
        \begin{alltt}
\element: \type{"V2X"} \{
  \field{"License"} IN
    [\val{"open source"}, \val{"third party"}] &
  \field{"Updates"} NOT IN [\val{"remote"}, \val{"yes"}]
\}\end{alltt}
        &
        \textbf{Targets:}\newline
        Integrity,\newline
        Availability,\newline
        Authorization\xrule{7.5em}\newline
        \textbf{Requires:} None
        \\
        \midrule
        $\upgamma_{13}$ &
        \begin{alltt}
\element: \type{"V2X"} \{
  \field{"Updates"} IN [\val{"remote"}, \val{"yes"}] &
  \field{"Secure Update"} != \val{"yes"}
\}\end{alltt}
        &
        \textbf{Targets:}
        Integrity,\newline
        Authorization,\newline
        Availability\xrule{7.5em}\newline
        \textbf{Requires:} None
        \\\bottomrule
    \end{tabular}
\end{table}

\begin{table}[t!]
    \centering
    \footnotesize
    \caption{Anti-Patterns for the Gateway asset $\asset_{\text{gateway}}$}
    \label{table:gateway}
    \begin{tabular}{p{1.65em}p{16.25em}p{9.25em}}
        \toprule
        \textbf{ID} & \textbf{Pattern} & \textbf{Comment} \\
        \midrule
        $\upgamma_{16}$ &
\begin{alltt}
\element: \type{"Gateway"} \{
    \field{"Authorization"} != \val{"yes"} |
    \field{"Data security integrity"} NOT IN
      [\val{"encrypted"}, \val{"secure hash"}]
\}\end{alltt}
        &
        \textbf{Targets:}
        Integrity,\newline
        Authorization,\newline
        Non-repudiation\xrule{9.25em}\newline
        \textbf{Requires:} $\iface_\text{WiFi}, \iface_\text{V2X}$
        \\
        \midrule
        $\upgamma_{17}$ &
        \begin{alltt}
\element: \type{"Gateway"} \{
  \field{"License"} IN
    [\val{"open source"}, \val{"third party"}] &
  \field{"Updates"} NOT IN [\val{"remote"}, \val{"yes"}]
\}\end{alltt}
        &
        \textbf{Targets:}
        Integrity,\newline
        Availability,\newline
        Authorization\xrule{9.25em}\newline
        \textbf{Requires:} $\iface_\text{WiFi}, \iface_\text{V2X}$
        \\
        \midrule
        $\upgamma_{18}$ &
        \begin{alltt}
\element: \type{"Gateway"} \{
  \field{"Updates"} IN [\val{"remote"}, \val{"yes"}] &
  \field{"Secure Update"} != \val{"yes"}
\}\end{alltt}
        &
        \textbf{Targets:}\newline
        Integrity
        \xrule{9.25em}\newline
        \textbf{Requires:} $\iface_\text{WiFi}, \iface_\text{V2X}$
        \\
        \midrule
        $\upgamma_{19}$ &
        \begin{alltt}
\element: \type{"Gateway"} \{
  \field{"Intrusion Prevention"} != \val{"yes"} &
  \field{"Intrusion Detection"} != \val{"yes"}
\}\end{alltt}
        &
        \textbf{Targets:}
        Integrity,\newline
        Non-repudiation\xrule{9.25em}\newline
        \textbf{Requires:} $\iface_\text{WiFi}, \iface_\text{V2X}$
        \\
        \midrule
        $\upgamma_{20}$ &
        \begin{alltt}
\element: \type{"Gateway"} \{
    \field{"DoS Mitigation"} != \val{"yes"}
\}\end{alltt}
        &
        \textbf{Targets:}
        Availability\xrule{9.25em}\newline
        \textbf{Requires:} $\iface_\text{WiFi}, \iface_\text{V2X}$
        \\\bottomrule
    \end{tabular}
\end{table}

\paragraph{Gateway:} A router acts as a gateway by applying intrusion prevention (IPS) and intrusion detection (IDS) to block any malicious incoming traffic from the WiFi access point ($\iface_{\text{WiFi}}$) and the V2X interface ($\iface_{\text{V2X}}$).
Thus, if a remote attacker attempts to exploit a gateway, they must first control $\iface_{\text{WiFi}}$ or $\iface_{\text{V2X}}$.
Therefore, each anti-pattern contains the $\iface_{\text{WiFi}}$ and $\iface_{\text{V2X}}$ interfaces to mark this relation.
\Cref{table:gateway} shows the resulting anti-patterns.
Finally, firewalls and IPS/IDS systems need DoS protection to counter a flooding attack.
In a flooding attack, an attacker creates a high volume of traffic that the firewall cannot handle.
If not protected, the firewall disables itself and lets all the data through.

\begin{table}[t!]
    \centering
    \footnotesize
    \caption{Anti-Patterns for the High-Level Controller asset $\asset_{\text{hlc}}$}
    \label{table:hlc}
    \begin{tabular}{p{1.65em}p{17.5em}p{8em}}
        \toprule
        \textbf{ID} & \textbf{Pattern} & \textbf{Comment} \\
        \midrule
        $\upgamma_{26}$ &
\begin{alltt}
\element: \type{"High-Level Controller"} \{
  \field{"License"} IN [\val{"closed source"},
               \val{"company internal"}]
& (\field{"Input Sanitization"} != \val{"yes"}|
   \field{"Coding guideline"}   != \val{"yes"}|
   \field{"Input Validation"}   != \val{"yes"})
\}\end{alltt}
        &
        \textbf{Targets:}\newline
        Integrity\xrule{8em}\newline
        \textbf{Requires:}\newline
        $\iface_\text{sensor}$,\newline
        $\asset_\text{gateway}$
        \\
        \midrule
        $\upgamma_{27}$ &
        \begin{alltt}
\element: \type{"High-Level Controller"} \{
  \field{"Updates"} IN [\val{"yes"}, \val{"remote"}] &
  \field{"Secure Update"} != \val{"yes"}
\}\end{alltt}
        &
        \textbf{Targets:}\newline
        Integrity\xrule{8em}\newline
        \textbf{Requires:} $\asset_\text{gateway}$
        \\
        \midrule
        $\upgamma_{28}$ &
        \begin{alltt}
\element: \type{"High-Level Controller"} \{
  \field{"License"} IN
    [\val{"open source"}, \val{"third party"}] &
  \field{"Updates"} NOT IN [\val{"remote"}, \val{"yes"}]
\}\end{alltt}
        &
        \textbf{Targets:}\newline
        Integrity,\newline
        Availability,\newline
        Authorization\xrule{8em}\newline
        \textbf{Requires:} $\asset_\text{gateway}$
        \\
        \midrule
        $\upgamma_{29}$ &
        \begin{alltt}
\element: \type{"High-Level Controller"} \{
  \field{"Service Name"} = \val{"network service"} &
  \field{"Privilege Separation"} != \val{"yes"}
\}\end{alltt}
        &
        \textbf{Targets:}\newline
        Authorization
        \xrule{8em}\newline
        \textbf{Requires:} $\asset_\text{gateway}$
        \\\bottomrule
    \end{tabular}
\end{table}

\paragraph{High-level controller ECU:} The final element that we include in this section is the High-Level Controller (HLC) ECU.
We consider it the target asset of the attacks.
For brevity, we only analyze this asset on a component level to keep the anti-pattern identification generic.

The HLC has different tasks, \eg, parsing sensor data, calculating the route, or sending commands to the Low-Level Controller.
\Cref{table:hlc} shows the resulting anti-patterns to each of the above tasks.
Of note, the HLC's sub-component level analysis is comprised of software components; therefore, we create anti-patterns that check the update procedures for third-party software and the coding guidelines for internal software.
These patterns depend on different entry points.
For example, the software parses either data from sensors ($\iface_{\text{sensor}}$) or data it receives through external communication with $\iface_{\text{WiFi}}$ and $\iface_{\text{V2X}}$.

Through similar analysis, we identified more interfaces, and component level and sub-component level assets; \eg the switch asset $\asset_\text{switch}$ and the Low-Level Controller asset $\asset_\text{llc}$.
For brevity, we disregard the identified anti-patterns for the remaining interfaces and asset\footnote{Note that disregarding the remaining assets and interface affects the continuity of anti-patterns' identifiers (\ie $\upgamma_n$) in the rest of this manuscript.}.

\subsubsection{Anti-Pattern Trees}
In the previous section, we collected relevant assets and anti-patterns in tables with which we build anti-pattern trees per asset.
We begin by transforming each table into an anti-pattern tree.
Then we identify a target asset and merge the trees.
The result of this section is a merged tree for a high-priority asset.

\paragraph{WiFi Access Point and Sensors:}
We presented the construction of the the anti-pattern trees for $\iface_{\text{WiFi}}$ and $\iface_{\text{sensor}}$ in \cref{fig:main:ats:wifi,fig:main:ats:sensor}, respectively.

\paragraph{V2X Interface:}
\Cref{fig:trees:v2x} illustrates the anti-pattern tree $\uptau_\text{V2X}$.
Since all the anti-patterns in \cref{table:Av2x} are independent, we connect them with an \OR node.
This tree represents an example, that can be extended to all identified interfaces and assets with no exploitability requirements.
Constructed anti-pattern sub-trees are then used for building the final anti-patterns tree that represents the vehicle's architecture.

\begin{figure}[!t]
  \centering
  \includegraphics[page=17,scale=0.75]{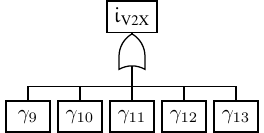}
  \caption{Anti-pattern tree $\uptau_\text{V2X}$ for the V2X interface $\iface_\text{V2X}$}
  \label{fig:trees:v2x}
\end{figure}

\paragraph{Gateway:}
The gateway asset $\asset_\text{gateway}$ represents an interface or asset with sequential exploitability requirements.
That is, to exploit a gateway asset, attackers must first control $\iface_\text{WiFi}$ or $\iface_\text{V2X}$.
Note \cref{table:gateway}.
To construct the anti-pattern sub-tree for such assets we must first build a set of requirement sub-trees that merge the exploitability requirements with correct root nodes.
Next, we connect requirements sub-trees to each anti-pattern in the \cref{table:gateway} using \SAND for sequential requirements and \AND otherwise.
See \cref{fig:gateway:extended}.

All sub-tree requirements for the gateway asset in \cref{fig:gateway:extended} are identical; \ie $\mathfrak{r}_1$ to $\mathfrak{r}_5$.
Hence, we can reduce the tree by grouping anti-patterns by requirements and connecting each anti-pattern group using an \OR node.
Thereafter, we connect each requirement sub-tree and its anti-pattern group using \SAND.
Finally, we combine all the resulting \SAND nodes using a root-level \OR node corresponding respective asset.
In case of $\mathfrak{r}_1$ to $\mathfrak{r}_5$ in \cref{fig:gateway:extended}, we represent them using a single sub-tree $\mathfrak{r}$.
Consequently, we combine the gateway's anti-patterns using a single \OR node; that is, $\uptau_\text{gateway}$ as shown in \cref{fig:gateway:reduced}.
Subsequently, we connect $\mathfrak{r}$ and $\uptau_\text{gateway}$ with \SAND.

\begin{figure}[!t]
    \centering
    \begin{subfigure}[b]{0.55\columnwidth}
      \centering
      \includegraphics [page=18,scale=0.75]{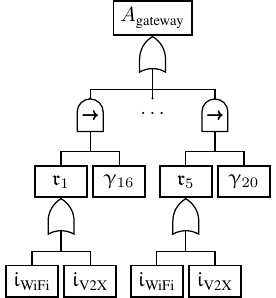}
      \caption{Extended anti-pattern tree}
      \label{fig:gateway:extended}
    \end{subfigure}%
    \begin{subfigure}[b]{0.45\columnwidth}
      \centering
      \includegraphics [page=19,scale=0.75]{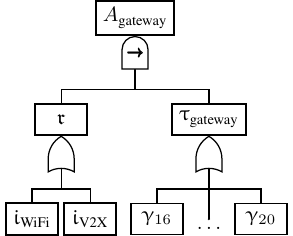}
      \caption{Reduced anti-pattern tree}
      \label{fig:gateway:reduced}
    \end{subfigure}
    \caption{Anti-Patterns trees for the gateway asset $\asset_\text{gateway}$}
    \label{fig:gateway}
\end{figure}

\paragraph{High-Level Controller:}
Since $\uptau_\text{hlc}$ in \cref{fig:hlc:abstract} is reasonably sized and contains some entry points, we showcase its construction.
The anti-pattern $\upgamma_{26}$ ensures all closed-source software sub-components within the HLC never jeopardize HLC's integrity.
An attacker can exploit any such sub-components that overlook input sanitization or validation or deviates from coding guidelines should (s)he also can exploit the gateway asset or the sensory interface.
Thus, $\upgamma_{26}$ requires $\iface_\text{sensor}$ or $\asset_\text{gateway}$ though not sequentially for such sub-component is already exploitable.
The rest of demonstrated patterns in \cref{table:hlc} only require $\asset_\text{gateway}$ sequentially for an attacker must implement these patterns through the gateway asset.
Note that we can group $\upgamma_{27}$ to $\upgamma_{29}$ with an \OR node as they share requirements.
\Cref{fig:hlc:merged} depicts the merged $\uptau_\text{hlc}$.
Finally, we can see infrastructure and sensory interfaces are the only entry points of the $\uptau_\text{hlc}$ and only in the higher levels of the tree we can gain access to inner components like the HLC.

\begin{figure}[!t]
  \centering
  \begin{subfigure}{\columnwidth}
    \centering
    \includegraphics [page=20,scale=0.75]{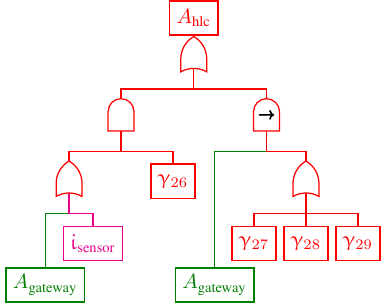}
    \caption{Abstract anti-pattern tree $\uptau_\text{hlc}$ for the HLC asset $\asset_\text{hlc}$}
    \label{fig:hlc:abstract}
    \vspace{2mm}
  \end{subfigure}
  \begin{subfigure}{\columnwidth}
    \centering
    \includegraphics [page=21,scale=0.75]{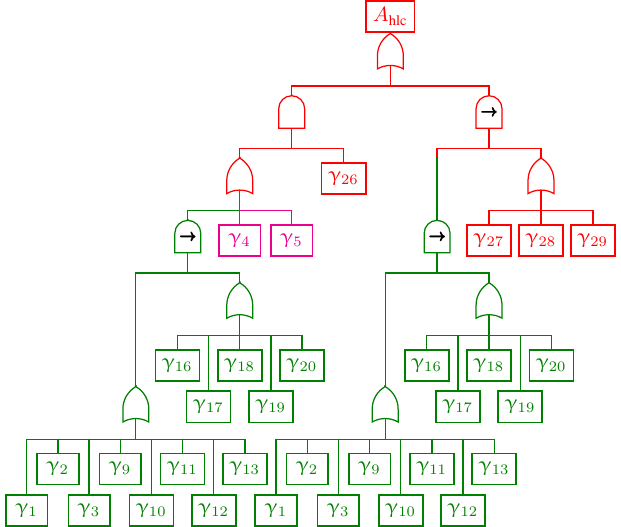}
    \caption{Merged anti-pattern tree $\uptau_\text{hlc}$ for the HLC asset $\asset_\text{hlc}$}
    \label{fig:hlc:merged}
  \end{subfigure}
  \caption{Anti-patterns trees for $\asset_\text{hlc}$ following the hierarchy of \cref{fig:model:enisa:tree}.}
  \label{fig:hlc}
\end{figure}

\subsubsection{Path Interpretation}
The anti-pattern tree construction interconnects assets with its identified anti-patterns.
However, we need to build the attack vectors to translate them into threat rules (\eg see \cref{table:running:rule,table:hlc:rule}). Although identifying attack vectors requires a deep understanding of multiple attack methods and how to combine them, the first step is to perform a SP graph interpretation of the anti-pattern trees.
Without the anti-pattern trees, such interpretation can be rather complex and would require support from experts in specific domains.

The SP graph of the anti-pattern tree for $\asset_\text{gateway}$ results in 40 different paths to attack the gateway asset.
Similarly, we consider two paths to attack the $\iface_\text{sensor}$.
In other words, there are $80$ paths to attack the $\asset_\text{hlc}$ using $\upgamma_{26}$  through either the gateway or the sensory interface.
Indeed, we can enumerate $80$ paths to reach $\upgamma_{26}$ in $\uptau_\text{hlc}$ as shown in \cref{fig:hlc:abstract}.
Similarly, there are $120$ paths to attack the $\asset_\text{hlc}$ using $\upgamma_{27}$ to $\upgamma_{29}$ through the gateway.
In total, the simplified anti-pattern $\uptau_\text{hlc}$ represents $200$ attack paths spanning from the outermost segments (\ie entry points) to the innermost segments that we analyzed here (\ie Car ECUs, Processing and Decision Making Components).

\subsubsection{Compiling Threat Rules}
Since all of the $120$ paths in $\spgraph{\uptau_\text{hlc}}$ can compromise the HLC from the SPIDER's surface, all of them are valid attack vectors that we must catalog.
This results in a threat rule that ThreatGet can automatically verify.
To exemplify the catalog process, we showcase the $\upgamma_2 \cdot \upgamma_{19} \cdot \upgamma_{29}$ path.
This path specifies a scenario in which exists a network service running by the HLC that disregards privilege separation.
To exploit the above vulnerability $\upgamma_{29}$ an attacker must first gain access to the vehicle's network $\upgamma_{2}$ and then bypass the intrusion detection and prevention systems $\upgamma_{19}$.
A compromised HLC through $\upgamma_{29}$ can lead to major safety consequences for the attacker issue safety impacting commands to the LLC taking over the actuators' control mechanism.

The attack path is a concrete description of the above SP graph path.
To this end, we now speculate how an attacker can realize $\upgamma_2 \cdot \upgamma_{19} \cdot \upgamma_{29}$.
The attacker can use a brute-force attack to implement $\upgamma_{2}$.
Next, (s)he can start injecting messages to the network service running on the HLC bypassing the IPS/IDS components.
Subsequently, (s)he can exploit the lack of privilege separation by implementing a privilege escalation attack for the network service.

Finally, the attacker must be within the WiFi range, the attack vector $V$ is adjacent.
The attack complexity $C$ is high, while the required privileges and user interaction are none.
\Cref{table:hlc:rule} specifies the threat rule for the explained threat scenario.

\begin{table}[!t]
    \footnotesize
    \centering
    \caption{Threat-Rule for the SP graph $\upgamma_2\cdot\upgamma_{19}\cdot\upgamma_{29}$}
    \label{table:hlc:rule}
    \begin{tabular}{p{5em}p{23.5em}}
        \toprule
        Title: &
        Compromise HLC through Network Service and WiFi
        \\
        \midrule
        Asset: & High-Level Controller -- Realtime Operating System\\
        \midrule
        Damage\newline
        Scenario: &
        A compromised HLC issues safety-impacting directives to the LLC controlling the vehicular actuators.
        \\
        \midrule
        Attack\newline
        Path: &
        1.\,Attacker brute-forces the WPA2 PSK.\newline
        2.\,Attacker injects messages in the network service that\newline remain undetected to the IPS/IDS system.\newline
        3.\,Attacker abuses the lack of privilege separation to elevate from network services to system applications.
        \\
        \midrule
        Threat\newline
        Scenario: &
        Failing to protect communication channels and\newline lacking privilege separation.
        \\
        \midrule
        Threat~Type: & Tampering \\
        \midrule
        Feasibility: &
        Medium ($V\!\!=\!\textit{adjacent}, C\!\!=\!\textit{high}, P\!\!=\!\textit{none}, U\!\!=\!\textit{none}$)
        \\
        \midrule
        Impact: & Major -- Safety \\
        \midrule
        Rule: &
\begin{alltt}
\flow \{
  SOURCE \element: \type{"WiFi Interface"} \{
    \field{"Key Length"} != \val{"long"}
  \} &
  TARGET \element: \type{"High-Level Controller"} \{
    \field{"Service Name"} == \val{"network service"} &
    \field{"Privilege Separation"} != \val{"yes"}
  \}
  INCLUDES \element: \type{"Gateway"} \{
    \field{"Intrusion Prevention"} != \val{"yes"} &
    \field{"Intrusion Detection"} != \val{"yes"}
  \}
\}\end{alltt}\\
\bottomrule
\end{tabular}
\end{table}
\section{Conclusion}
\label{sections:conclusion}
As required by \un and aiming at the \iso work products, automotive cybersecurity analysis introduces multiple challenges.
Besides the obvious, like repeatability and completeness of the analysis, an iteration over all assets and attack surfaces and resulting paths generates many artifacts that need to be considered and further pursued throughout the complete engineering process.

We proposed a methodology considering the above requirements.
We first model the system using a DFD and formalize its cybersecurity requirements using attack trees representing anti-patterns for cybersecurity requirements.
We define model refinement through systematic segmentation of the vehicle.
Moreover, the proposed segmentation could follow a community guideline like the ENISA Smar Car hierarchical reference model.
We showed how to extend segmentation to the formalization step and construct sub-trees specifying an attack per segment.
Afterward, we extensively described merging these sub-trees into anti-pattern trees to perform system-level cybersecurity analysis.
With the generation of attack trees and the formalization of the resulting threat understanding in reusable rules, we show how complexity can be managed and ease the challenge of repeated analysis.
Our proposed methodology can be mapped to the work products and steps required in \iso and is therefore compliant with the requirements on a TARA for automotive systems.
We demonstrated that this method could handle real-world-sized systems by applying it to a real-world use case.

Finally, we argue that although we created a new formalism for attack trees, applying the same principles to existing attack trees is still possible.
We can also translate attack trees consisting of goals, sub-goals, and attack steps into our formalism by replacing goals with assets, sub-goals with entry points, and attack steps with anti-patterns.
The main task is to translate the content of nodes (\eg, written statements or logic propositions) into ThreatGet's DSL, similar to what we show in \cref{sections:methodology:formalize}.
Then, by applying the following steps (path interpretation and compiling threat rules), we reuse the knowledge collected in those trees and model it formally so that ThreatGet can interpret it.
While the initial effort is higher than the manual analysis, this effort also allows for documenting threat knowledge from different specializations (cryptography, Hardware-security, Wireless Attacks, \etc) and generates a reusable set, reducing future efforts.
\section{Acknowledgments}
We would like to thank Roderick Bloem for his valuable comments and suggestions.
\wwgrant 
The Virtual Vehicle Research GmbH would also like to acknowledge the financial support within the COMET K2 Competence Centers for Excellent Technologies from the Austrian Federal Ministry for Climate Action (BMK), the Austrian Federal Ministry for Digital and Economic Affairs (BMDW), the Province of Styria (Dept. 12) and the Styrian Business Promotion Agency (SFG). The Austrian Research Promotion Agency (FFG) has been authorised for the programme management.

\balance
\setlength{\bibsep}{0.0pt}
\bibliographystyle{ieeetr} 
\bibliography{IEEEabrv,references.bib}

\vspace{2pt}
\section{Bib\TeX~Entry for Citation}
\begin{verbatim}
@inproceedings{SAEThreatModeling2022,
  title        = {Identification and Verification of Attack-
                  Tree Threat Models in Connected Vehicles},
  author       = {Masoud Ebrahimi and Christoph Striessnig and
                  Joaquim Castella Triginer and
                  Christoph Schmittner},
  booktitle    = {SAE Technical Paper 2022-01-7087},
  doi          = {10.4271/2022-01-7087},
  year         = {2022}
}
\end{verbatim}

\end{document}